\documentstyle[11pt,aspconf]{article}

\begin{document}

\title{The Tucana dwarf galaxy}
\author{M. Castellani $^{1,2}$}
\author {G. Marconi$^2$}
\author {R. Buonanno$^2$}
\affil{(1) Universita' ``La Sapienza'', Istituto Astronomico, Via Lancisi 29,
00161 Roma, Italy}
\affil{(2) Osservatorio Astronomico di Roma, Via dell'Osservatorio 2, 
Monte Porzio Catone, 00040 (Roma), Italy}

\begin{abstract}
Deep CCD photometry for the dwarf galaxy in Tucana is presented.
Distance modulus and metallicity are derived, together with an estimate of
the age of the galaxy.

\end{abstract}


\section{Introduction}
We present deep CCD photometry for the dwarf galaxy in Tucana
(l=323, b=-47.4). The observations were made using the high
resolution camera SUSY on the 3.5 m NTT of ESO - La Silla in the
period from November 4 to November 7 1994.
 A coated Tektronix tk 1024 X 1024 pixels CCD was used at the Nasmith
 focus A, with a scale of 0.12'' per pixel. Several frames of a field  
centered on the main body ($\alpha_{1950}
$=$22^{h}38^{m}27.9^{s}$, $\delta_{1950}$=$-0.64^{0}$40'53'')
of the galaxy were obtained. 
The observed
area corresponds roughly to the 80$\%$ 
of the luminous main body of the galaxy.
We also show a preliminary CM diagram of Tucana obtained from HST 
observations, taken at ESO archive.

\section{CM diagram}

Fig.1a shows the V vs V-I diagram of the stars obtained after the reduction
procedures. 
Fig.1b shows the same CM diagram obtained with data taken at HST
archive. In this last figure is easily seen the signature 
of an horizontal branch. 
Red stars apparently describe a well-defined sequence in the CM diagram very
similar to the ``enlarged RGB'' shown by 
other dwarf spheroidals, like Carina (Smecker-Hane et al. 1994) or Phoenix
(Ortolani and Gratton 1988).
\begin{figure}
\plotfiddle{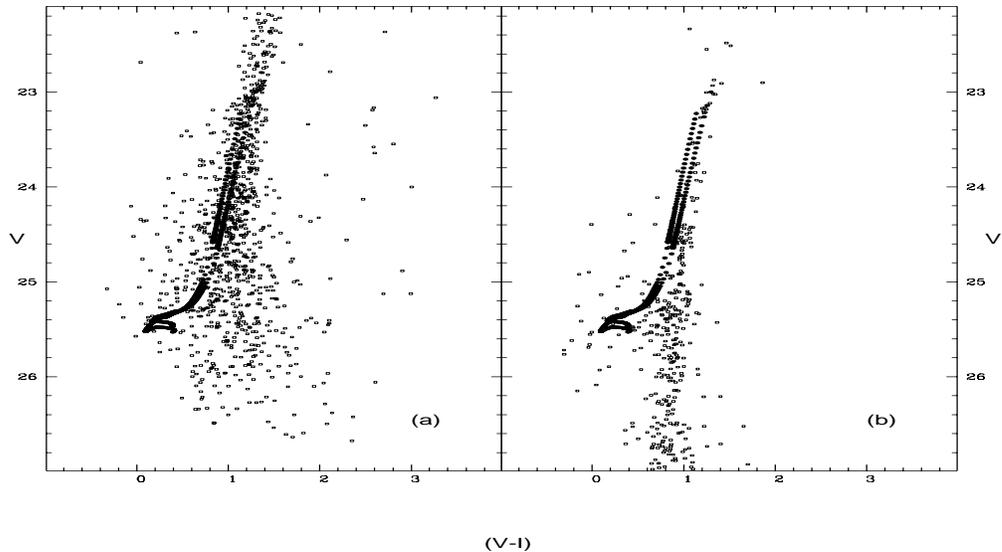}{8truecm}{0}{70}{40}{-210}{-50}
\caption{CM diagrams of Tucana: (a) NTT data (b) HST data}
\end{figure}

Please note that 
the spread in colour is larger than the one expected 
on the basis of photometric errors.
This could indicate that Tucana is similar, in this respect,
to the dwarf spheroidal Carina, and that
part of this increase is due to 
a superimposed AGB population.
On the other hand, 
Tucana shares with the ``transition'' dwarf galaxy in Phoenix 
the complete lack of blue bright main
sequence (MS) stars, i.e. 
bright stars with V-I $\simeq$ 0.
This suggests that {\it Tucana is composed
mainly of an old stellar population.}

\section{Distance modulus and metallicity}

If we interpret the sequence of Fig.1 as the RGB of an
intermediate-old stellar population,
we can use  
the location  
and the shape 
of such RGB to estimate the {\it distance} of the dwarf galaxy in Tucana,
through the comparision 
with the giant branch loci of a sample of galactic globular clusters.

We therefore computed the ridge line of the RGB of Tucana by  
spline fitting and then, given the negligible reddening of the
region of Tucana (see the maps of Burstein and Heiles 1982),
we shifted this ridge line vertically to obtain the best match with the
sample of globular clusters given by 
Da Costa and Armandroff (1990). 
This sample ranges in metallicity  
from [Fe/H]=-0.71 to
[Fe/H]=-2.17.

Using this procedure, we find that 
the locus of the Tucana giant-branch appears
to overlap on the giant branch
loci 
of NGC6752 ([Fe/H]=-1.54) and of M2 ([Fe/H]=-1.58), {\it for
distance modulus $(m-M)_I=24.72\pm0.20$}, 
as shown in Fig.2.
This agrees with a previous estimate
of $(m-M)_V=24.8\pm0.2$ obtained by Da Costa (1994).

\begin{figure}
\plotfiddle{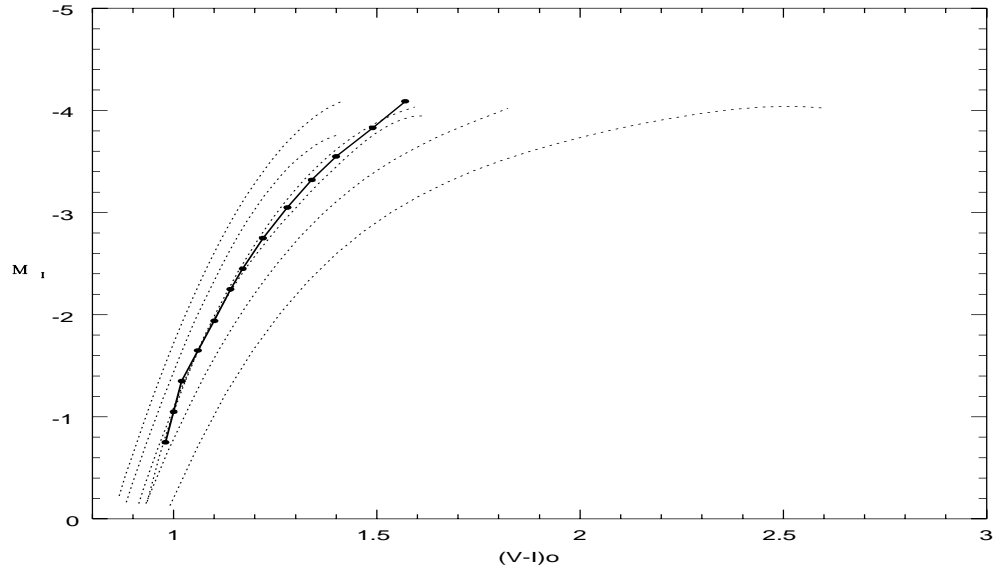}{6.5truecm}{0}{70}{40}{-210}{-50}
\caption{Giant branch loci of six globular cluster (from left, of M15,
NGC 6397, M2, NGC 6752, NGC 1851, and 47 Tuc) (from Da Costa and 
Armandroff 1990) and of Tucana (bold line, NTT data)}.
\end{figure}

{\it We will therefore assume [Fe/H]=-1.56$\pm$0.20 as the metallicity 
of Tucana.} This figure is consistent with the suggestion of
Gallagher and Wise (1994), that most dwarf galaxies have mean 
metallicities near to the peack of metallicities of the halo
globular clusters.

Finally, the distance modulus derived above has been
nicely confirmed by a different procedure, based on the {\it use of 
AGB clump as
a standard candle} (Pulone 1992). In Fig.1 the AGB
evolutionary tracks, for two models of Z=0.0004 and Z=0.001, Y=0.23, M=0.7
$M_{\odot}$ (Castellani, Chieffi \& Pulone 1991) are also shown, 
scaled to the CMD of Tucana using
the obtained distance modulus. 

\section{Age estimate}

One finds hints that
{\it Tucana is composed, almost
exclusively, of an intermediate-old stellar 
population} (t$\geq$5 Gyr).  
This is supported by the lack of blu MS supergiants and red
supergiants.

Moreover, the existence of stars of intermediate age and 
more massive than 1.5$M_{\odot}$ would have been revealed by the presence 
of a TO structure (in Fig.1) in the region $V\leq24$, $(V-I)\simeq0$.

\begin{figure}
\plotfiddle{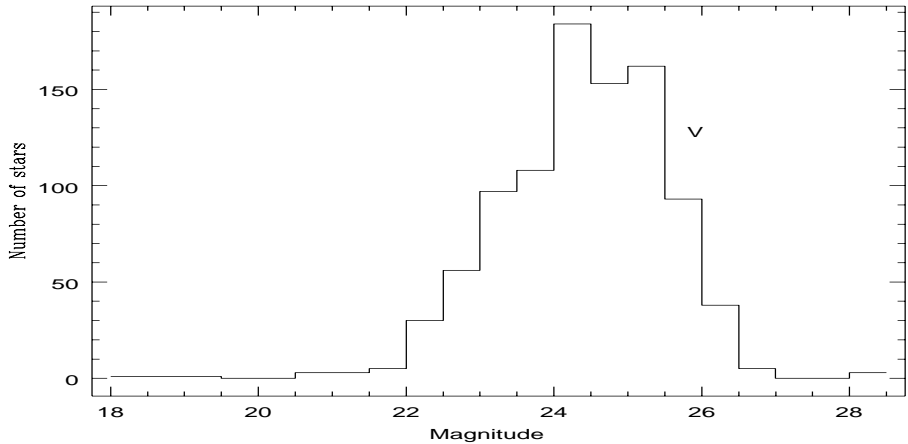}{6.5truecm}{0}{70}{40}{-210}{-50}
\caption{Luminosity function of Tucana (in V band, NTT data)}
\end{figure}

In the light of this conclusion, 
the bump observed in the luminosity function at V $\simeq$
24.3 mag (see Fig.3) is likely to be interpreted 
as {\it the signature of the AGB clump}, due 
to the slowing down in the
evolutionary rate, preceded by a fast evolution at the end of the 
core He-burning phase. This is well in agreement with the
magnitude of the horizontal branch, V=25.5, as easily desumed from Fig.1b.
We note as the magnitude of ZAHB of the overplotted tracks nicely fits the  
locus of maximum density of stars in CM diagram.

\section{Discussion and summary}

We have obtained V, I CCD photometry (NTT, HST) for the dwarf galaxy 
in Tucana, down to the limit of V$\simeq$27.5.

The {\bf distance modulus} has been derived by different procedures,
all converging to a value of $(m-M)_V=24.7\pm0.2$ 

The {\bf metallicity} have been derived by comparing the RGB locus 
of Tucana to the ridge-lines of globular clusters of
different metallicities, thus
obtaining a value of $[Fe/H]\simeq-1.56$. 
 
We suggest that
the CM diagrams we obtained allow a strightforward interpretation
in terms of {\it old low-mass stars} in the phase of H-burning in shell (RGB).

In the HST CM diagram (Fig.1b), we also detected {\it a clear signature of
an horizontal branch.}

We did not detect hints of 
recent and conspicuons episodes of star formation.

The intrinsic dispersion along the RGB has been
interpreted as {\it a dispersion in metallicity of the stars in Tucana},
although the presence of an intermediate population of few AGB
stars (M $\simeq$ 2M$_{\odot}$) 
cannot be excluded.


\end{document}